\documentclass[epj]{svjour}
\usepackage{latexsym}
\usepackage{graphics}
\usepackage{amsmath}
\usepackage{amssymb}
\usepackage{bbm}
\usepackage{epsfig}
\newcommand{\Eqref}[1]{Eq.~\eqref{#1}}
\begin{document}
\title{Strong laser fields as a probe for fundamental physics}
\author{Holger Gies
}                     
\institute{Theoretisch-Physikalisches Institut,
  Friedrich-Schiller-Universit\"at Jena, Max-Wien-Platz 1, D-07743 Jena, Germany}
\date{Received: date / Revised version: date}
%
\abstract{ Upcoming high-intensity laser systems will be able to probe the
  quantum-induced nonlinear regime of electrodynamics. So far unobserved QED
  phenomena such as the discovery of a nonlinear response of the quantum
  vacuum to macroscopic electromagnetic fields can become accessible. In
  addition, such laser systems provide for a flexible tool for
  investigating fundamental physics. Primary goals consist in verifying so far
  unobserved QED phenomena. Moreover, strong-field experiments can search for
  new light but weakly interacting degrees of freedom and are thus
  complementary to accelerator-driven experiments. I review recent
  developments in this field, focusing on photon experiments in strong
  electromagnetic fields. The interaction of particle-physics candidates with
  photons and external fields can be parameterized by low-energy effective
  actions and typically predict characteristic optical signatures. I perform
  first estimates of the accessible new-physics parameter space of
  high-intensity laser facilities such as POLARIS and ELI.
%
} 
\maketitle
\section{Introduction}
\label{intro}

The superposition principle of classical electrodynamics is violated on the
quantum level. Heisenberg's uncertainty principle allows for fluctuations of
electron-positron pairs on top of the vacuum. As these fluctuations separate
char\-ges on a time and length scale on the order of the Compton wavelength,
electromagnetic fields can couple to the charge fluctuations. On average,
these vacuum-mediated interactions induce nonlinearities among the
electromagnetic field itself, giving rise to nonlinear corrections to
Maxwell's theory \cite{Heisenberg:1935qt,Weisskopf:1936,Schwinger:1951nm}. 

On a microscopic level, fluctuating electrons and posi\-trons and their
interaction with electromagnetic fields is described by quantum
electrodynamics (QED). QED has been tested to an extremely high precision in
atomic physics and at accelerator experiments. By contrast, the predicted
violation of the superposition principle for macroscopic classical fields has
not been verified so far. Sizable deviations from linearity require field
strengths on the order of the energy scale set by the mass of the fluctuating
particle, namely the electron mass: $B_{\text{cr}}= m^2/e \simeq 4.3\times
10^{9}$Tesla or $E_{\text{cr}}\equiv B_{\text{cr}}=1.3\times
10^{18}$Volt/meter (we use units such that $\hbar=c=1$). This exceeds standard
laboratory electromagnetic field strengths by far. For such fields with $E,B
\ll B_{\text{cr}}$, the nonlinear interactions to lowest order are governed by
the Heisenberg-Euler effective action, $\Gamma_{\text{HE}}= \int d^4x \,
\mathcal{L}_{\text{HE}}$.  The corresponding effective Lagrangian is given by
\cite{Heisenberg:1935qt,Weisskopf:1936,Schwinger:1951nm}
\begin{equation}
  \mathcal{L}_{\mathrm{HE}} = -\mathcal{F}+ \frac{8}{45}\, \frac{\alpha^2}{m^4} \,
  \mathcal{F}^2 + \frac{14}{45}\, \frac{\alpha^2}{m^4}\, \mathcal{G}^2, 
\label{eq:HE}
\end{equation}
where $\mathcal{F} = \frac{1}{4} F_{\mu\nu}F^{\mu\nu}=\frac{1}{2}
(\mathbf{E}^2 - \mathbf{B}^2)/2$, $\mathcal{G} = \frac{1}{4}
F_{\mu\nu}\widetilde{F}^{\mu\nu}=\mathbf{E} \cdot \mathbf{B}$. In addition to
the Maxwell part, $\mathcal{L}_{\text{M}}=-\mathcal{F}$, the four-photon
interaction terms $\sim \mathcal{F}^2,\mathcal{G}^2$ arise from the average
over the electron-positron fluctuations, see
\cite{Dittrich:2000zu,Dunne:2004nc} for reviews. Apart from the suppression by
the electron mass scale, nonlinear phenomena are also suppressed by factors of
$\alpha\simeq 1/137$.

In order to observe the nonlinearities for weak fields in the laboratory, long
interaction lengths or times are required. This scheme has been pioneered by
the BFRT experiment \cite{Cameron:1993mr} and more recently by PVLAS
\cite{Zavattini:2007ee}. Here, an interaction of optical laser photons with
magnetic fields of order $\mathcal{O}(1-10\text{T})$ is searched for using
cavity techniques in order to increase the optical path length up to
$\mathcal{O}(10\text{km})$. Even though QED nonlinearities are still a few
orders of magnitude below the current sensitivity scale, these experiments
have already probed a new window of particle physics, see below. 

By contrast, upcoming high-intensity laser systems have the potential to get a
more direct access to strong-field nonlinearities. Whereas the field intensity
parameter $\nu=B^2/B_{\text{cr}}^2\sim 10^{-18}$ for BFRT or PVLAS, current
Peta-Watt lasers such as POLARIS at Jena \cite{POLARIS} will reach up to
$\nu\sim 10^{-7}$. At planned facilities such as ELI \cite{ELI}, a maximum of
$\nu\sim 10^{-3}$ is expected. Since this drastic enhancement of field
strengths also requires a strong focusing of the laser pulse, the possible
interaction region is only on the order of
$\mathcal{O}(10-100\mu\text{m})$. A discussion of the benefits and
disadvantages of laser-based experiments will be given in this article. 

The discovery of the quantum-induced nonlinearities of electrodynamics would
verify our most successful theory QED in a parameter region which has been
little explored so far. It would complete a quest which has begun in the
thirties. But beyond this, there is another strong motivation to investigate
strong-field nonlinearities, since these experiments have a discovery
potential of new fundamental physics. This is because the source of
fluctuation-induced nonlinear self-interactions of strong electromagnetic
fields in vacuum is not restricted to electrons and positrons. Any quantum
degree of freedom that couples to photons can lead to modifications of
Maxwell's electrodynamics. Strong-field experiments therefore also investigate
the general field content of fluctuating particles in the quantum vacuum. If
so far unknown particles mediate apparent photon self-interactions in a manner
similar to elec\-tron\--positron fluctuations, they can generate nonlinear
corrections analogous to the Heisenberg-Euler Lagrangian
\eqref{eq:HE}. Moreover, such hypothetical particles could even be created
from strong fields. As high-intensity lasers such as POLARIS and ELI will
substantially push the frontier of strong fields available in a laboratory,
they have the potential to search directly and indirectly for new fundamental
particles.

As the scale set, for instance, by the ELI peak field strength is expected to
be of order $\mathcal{O}(100\text{keV})$ in particle-mass/energy units, a
typical strong-field experiment will be sensitive to particle masses up to
this scale and particularly to much lower scales. On the other hand, the high
field strength together with modern optical techniques provides for a strong
handle on very weakly coupled particles. With this particular sensitivity to
potentially light but weakly coupled degrees of freedom, strong-field
experiments are complementary to accelerator searches for new particles
\cite{Gies:2007ua}.

Indeed a number of extensions of the Standard Model of particle physics
predict the existence of weakly interacting sub-eV particles (WISPs) which
couple to the electromagnetic sector. A popular candidate is the axion
\cite{Peccei:1977hh} which provides for a possible solution of the strong CP
problem; more generally, we can think of axion-like particles (ALPs) as an
uncharged scalar or pseudo-scalar degree of freedom with a coupling to two
photons. Further candidates are mini-charged particles (MCPs), i.e., matter
fields with charge $\epsilon e$ and $\epsilon \ll 1$, which arise naturally in
scenarios with gauge-kinetic mixing \cite{Holdom:1985ag} or extra-dimensional
scenarios \cite{Batell:2005wa}. More generally, many Standard-Model extensions
not only involve but often require -- for reasons of consistency -- a hidden
sector, i.e., a set of so far unobserved degrees of freedom very weakly
coupled to the Standard Model. Hence, a discovery of hidden-sector properties
could decisively single out the relevant theoretical fundament.

In this article, we summarize both our well-founded QED expectations for
nonlinear phenomena as well as well-moti\-vated speculations of possible
signatures for new physics in strong-field experiments. By performing first
estimates of the parameter region accessible to high-intensity laser systems,
we will argue that such systems can become an important source of information
for fundamental physics both within and beyond the Standard Model of particle
physics.

\section{Low-energy effective actions}
\label{sec2}

From a bottom-up viewpoint, QED as well as many extensions of the Standard
Model of particle physics lead to similar consequences for low-energy
laboratory experiments. These are parameterizable by effective nonlinear
interactions such as \Eqref{eq:HE} or effective couplings between photons and
the new effective degrees of freedom. In the following, we summarize a set of
generic low-energy effective actions as well as typical observables for
optical experiments involving strong fields.

\subsection{Heisenberg-Euler effective action}

The first example is given by classic QED, effectively summarized by the
Heisenberg-Euler effective action \Eqref{eq:HE} governing the dynamics of
macroscopic laboratory fields. This action gives rise to a number of
non-classical phenomena, most prominently vacuum-electromagnetic birefringence
\cite{Toll:1952rq,Adler:1971wn,Dittrich:2000zu}. The quantum-modified
equations of motion resulting from \Eqref{eq:HE} yield
\begin{equation}
  0=\partial_\mu \left({F^{\mu\nu}} - \frac{4}{45}
    \frac{\alpha^2}{
      m^4}{F^{\alpha\beta}F_{\alpha\beta}} 
    {F^{\mu\nu}}
    - \frac{7}{45} \frac{\alpha^2}{ m^4}
    {F^{\alpha\beta}F_{\alpha\beta }} {\widetilde
      F^{\mu\nu}} \right) .\label{eq:ME}
\end{equation} 
In comparison to Maxwell's equation $\partial_\mu F^{\mu\nu}=0$, this equation
no longer admits plane wave solutions traveling at the speed of light. But
for a probe beam with small amplitude propagating in a strong background
field, we may linearize the field equation and solve for the dispersion
relation of the probe beam. For instance for a strong background magnetic
field $B$, the phase and group velocities of the probe field satisfy
\begin{equation}
  { v_\|}\simeq 1- \frac{14}{45} \frac{\alpha^2}{m^4}\,
   B^2 \sin^2 \theta_B, \quad 
  { v_\bot}\simeq 1- \frac{8}{45} \frac{\alpha^2}{m^4}\,
   B^2 \sin^2 \theta_B, \label{eq:velo}
\end{equation}
where $\theta_B$ is the angle between the propagation direction and the
magnetic field. The indices $\|$ and $\bot$ distinguish the two polarization
modes, where the polarization of the probe beam is in or perpendicular to the
plane spanned by the propagation direction and the $B$ field. 
For a probe field in a counter-propagating laser field,
$B^2\sin^2\theta_B$ has to be replaced by the laser intensity $I$
\cite{Heinzl:2006xc}.  

The vacuum modified by a strong external field thus is birefringent in a
manner similar to a uniaxial crystal with refractive indices
$n_{\|,\bot}=1/v_{\|,\bot}$.  As an observable, an initially linearly
polarized probe laser can pick up an {\em ellipticity} by traversing the
strong beam. The ellipticity angle $\psi$ is given by 
\begin{equation}
\psi=\frac{\omega}{2} L \Delta n\, \sin 2 \theta,\label{eq:ell}
\end{equation}
where $\theta$ is the angle between the probe polarization and the fast
eigenmode's polarization, $\Delta n = n_\|- n_\bot$, and $L$ is the path
length inside the strong field.

Equation \eqref{eq:ell} can teach a lot about the specific advantages and
disadvantages of the various possible setups. Classic strong-field experiments
such as BFRT or PVLAS at comparatively low field strengths need to detect a
tiny refractive-index difference on the order of $\Delta
n\simeq10^{-22}$. Using optical probe photons, e.g., $\omega \simeq 1\mu$m,
high-finesse interferometry can increase the optical path length to up to
$L\simeq \mathcal{O}(10\text{km})$. By contrast, high-intensity lasers such as
POLARIS can reach up to $\Delta n\simeq 10^{-12}$, but are restricted to
optical path lengths of order $L\simeq\mathcal{O}(10\mu\text{m})$ dictated by
focusing as close to the diffraction limit as possible. Since cavity
techniques are useless anyway for short-pulsed high-intensity fields, a higher
probe frequency $\omega$ can be used in order to enhance the ellipticity
signal, as proposed in \cite{Heinzl:2006xc}.

For instance, for x-ray photons with $\omega\simeq 1$keV probing a POLARIS
high-intensity field, the induced ellipticity angle is given by $\psi
\simeq 6 \times 10^{-7}$. Higher frequencies such as $\omega\simeq 12$keV
even yield $\psi \simeq 7 \times 10^{-6}$. This should be compared to
the sensitivity scale of an x-ray polarization measurement required for
detecting ellipticity. Theoretical estimates predict that sensitivity bounds on
the order of $\psi\simeq 3\times 10^{-6}$ should be measurable with
high-precision techniques for these frequencies \cite{Alp}. Therefore,
quantum-induced nonlinearities of electromagnetic fields may already be
discovered at Peta-Watt lasers such as POLARIS. Indeed, even the required
synchronized x-ray beam can be generated via laser-driven electron
acceleration and subsequent Thomson backscattering with the same system. 

Another optical observable can be important: any effect which modifies the
amplitudes of the $\|$ or $\bot$ components in a polarization-dependent manner
but leaves the phase relations invariant will induce a {\em rotation} angle
$\Delta \theta$. Since amplitude modifications involve an imaginary part for
the index of refraction, rotation from a microscopic viewpoint is related to
particle production or annihilation. In QED below threshold $\omega<2 m$,
electron-positron pair production by an incident laser is excluded.  Further
possibly rotation inducing effects such as photon splitting
\cite{Adler:1971wn} or neutrino-pair production \cite{Gies:2000wc} in a strong
field are severely suppressed for typical laboratory parameters.  Therefore, a
sizeable signal for vacuum rotation $\Delta \theta$ in a strong-field
experiment would be a signature for new fundamental physics.  (Note, however,
that rotation can also be generated for pure kinematical reasons, depending on
the details of the optical set up \cite{DiPiazza:2006pr}.)

Let us close this QED part by stressing, that we have concentrated on
nonlinear QED phenomena with external fields as sole asymptotic
states. Another very important and interesting nonlinear phenomenon is the
spontaneous decay of the electromagnetized vacuum itself into
electron-positron pairs. This Schwinger pair production is technically related
to the imaginary part of the effective action $\text{Im} \Gamma$
\cite{Heisenberg:1935qt,Weisskopf:1936,Schwinger:1951nm}. In addition to being
nonlinear, it is also nonperturbative in the coupling to the, say, electric
background field $eE$, see \cite{Dunne:2004nc,Dunne:thisvolume} for a
review. An experimental verification of this phenomenon would therefore
explore a particularly interesting and incompletely understood branch of
quantum field theory. For subcritical fields, the pair production rate is
unfortunately exponentially suppressed $\sim \exp(-\pi E_{\text{cr}}/E)$
according to a constant-field approximation. Time-dependent fields can enhance
the production rate significantly
\cite{Dunne:2004nc,Dunne:thisvolume,Dunne:1998ni}. Recent results indeed
indicate that th exponential suppression might be overcome with time-dependent
tailored pulses even for ELI parameters \cite{Schutzhold:2008pz}.

\subsection{Axion-Like Particle (ALP)}

As a first example of a new particle candidate beyond those degrees of freedom
of the Standard Model, we consider a new neutral scalar $\phi$ or pseudo-scalar
degree of freedom $\phi^{-}$ such as an axion which is coupled to the photon by,
\begin{equation}
\mathcal{L}_{\text{ALP}}= \left\{-\frac{{g}}{4} \phi^{(-)} F^{\mu\nu}
\stackrel{(\sim)}{F}_{\mu\nu}
-\frac{1}{2}(\partial\phi^{(-)})^2 - \frac{1}{2}  {m_\phi}^2 \phi^{(-)}{}^2
\right\},
\label{eq:ALP}
\end{equation}
parameterized by the mass $m_\phi$ of this axion-like particle (ALP) and the
dimensionful coupling $g$. In optical experiments in strong fields, ALPs can
induce both ellipticity and rotation \cite{Maiani:1986md}, since only one
polarization mode couples to the ALP and the strong field. For instance,
coherent photon-ALP conversion causes a depletion of one photon mode, implying
rotation. In order to make contact with the literature, we approximate the
strong field by a homogeneous magnetic field $B$ as may be provided by a
slowly beating standing wave formed from counter-propagating laser beams. We
stress that detailed studies employing all relevant properties of the field
provided by systems such as ELI still need to be performed.  From the
equations of motion for the photon-ALP system for the pseudo-scalar case, the
induced ellipticity and rotation can be calculated:
\begin{eqnarray}
\psi^{-}&
=& \frac{1}{2}\left({\scriptstyle
  \frac{gB\omega}{m_\phi^2}}\right)^2\left({\scriptstyle \frac{L m_\phi^2}{2\omega}}-
\sin\!\left({\scriptstyle \frac{L
    m_\phi^2}{2\omega}}\right)\right)\sin2\theta,
    \nonumber\\
{{\Delta\theta}}^{-}
  &=& \left({\scriptstyle \frac{gB\omega}{m_\phi^2}}\right)^2
    \sin^2\!\left({\scriptstyle \frac{L m_\phi^2}{4\omega}}\right)\sin2\theta,
\label{eq:ellrotALP}
\end{eqnarray} 
for single passes of a probe beam through a strong $B$ field of length $L$.
For the scalar ALP, we have $\Delta\theta=-\Delta\theta^{-}$,
$\psi=-\psi^{-}$. Measuring ellipticity and rotation signals uniquely
determines the two model parameters, ALP mass $m_\phi$ and ALP-photon coupling
$g$.  Measuring the signs of $\Delta \theta$ and $\psi$ can even resolve the
parity of the involved particle \cite{Ahlers:2006iz}.

The effective interaction \eqref{eq:ALP} is representative for various
underlying particle scenarios. In the axion case, only the weak coupling to
the photon is relevant and all other potential matter couplings are
negligible. This facilitates the interesting experimental option to shine the
ALP component through a wall which blocks all photons. Behind the wall, a
second strong field can induce the reverse process and photons can be
regenerated out of the ALP beam \cite{Sikivie:1983ip}. The regeneration rate
is
\begin{equation}
\label{regALPps}
n_{\text{out}}= n_{\text{in}}\,
\frac{1}{16}\left(gBL\cos\theta\right)^4\left[
{\sin\left(\frac{L m_\phi^2}{4\omega}\right)} \Big/
{\frac{L m_\phi^2}{4\omega}}
\right]^4,
\end{equation}
where $n_{\text{in}}$ is the initial photon rate, and the fields $B$ and its
extension $L$ are assumed to be identical on both sides of the wall.

Further models with ALPs have been proposed in cosmology. Cosmological scalar
fields are discussed, e.g., in the context of dark energy, the cosmological
coincidence problem and also dark-matter abundance. An interesting candidate
also for optical experiments are those scalar fields with a chameleon
mechanism which have been developed in the context of the fifth-force problem
\cite{cham}. As an interesting property, a chameleonic ALP cannot penetrate
the end caps of the vacuum chamber but gets reflected back. Whereas this has
no influence on the formulas for $\psi$ and $\Delta \theta$ in
\Eqref{eq:ellrotALP}, a new detection mechanism arises: synchronizing a short
laser probe pulse with the strong pulse, chameleons can be created inside the
vacuum chamber and stored in a parallel cavity. By a synchronized second
strong pulse, the chameleons can be re-converted into photons again inside the
strong field; this would result in an afterglow phenomenon which is
characteristic for a chameleonic ALP \cite{Gies:2007su}. In the parameter
range where $gB/m_\phi\ll 1$, the number of photons in the first afterglow
pulse $n_{\text{out}}$ is again given by \Eqref{regALPps} where this time
$n_{\text{in}}$ is the number of photons in the synchronized probe pulse
initially generating the chameleons.

\subsection{Minicharged Particle (MCP)}

In addition to neutral particles coupling to photons by dimensionful coupling
constants such as ALPs, new particle candidates can also be charged. In order
to have evaded detection so far, they have to be either very heavy or very
weakly charged. In the latter case, they can also be very light and thus
become ideal candidates for laser-based searches.  These so-called minicharged
particles (MCPs) arise naturally in scenarios with gauge-kinetic mixing
\cite{Holdom:1985ag} or extra-dimensional scenarios \cite{Batell:2005wa}, and
find a natural embedding in string-theory models with intermediate string
scale \cite{Abel:2006qt}. The latter property makes optical searches for MCPs
particularly attractive, because a possible optical signal at low energies
could already help singling out the relevant class of microscopic
highest-energy models. 

From the bottom-up viewpoint of effective photon interactions and
nonlinearities, the fluctuations of mini\-char\-ged particles with mass
$m_\epsilon$ and charge $\epsilon e$, $\epsilon \ll 1$ induce photon
self-interactions in the same way as electrons do. However, the weak-field
Heisenberg-Euler Lagrangian \eqref{eq:HE} is not sufficient to describe the
physics of MCPs properly, as the expansion parameters $\epsilon e
B/m_\epsilon^2$ and $\omega/m_\epsilon$ in a strong $B$ field varying with
$\omega$ are not necessarily small. As an interesting consequence, the probe
laser frequency can be above the pair-production threshold
$\omega>2m_\epsilon$ such that a rotation signal in addition to
birefringence-induced ellipticity becomes possible \cite{Gies:2006ca}.

All relevant information is encoded in the polarization tensor which is well
known from QED \cite{Toll:1952rq,Daugherty:1984tr,Dittrich:2000zu}. Explicit
results are available in asymptotic limits, e.g., for the rotation signal
induced by a Dirac-fermionic MCP \cite{Gies:2006ca},
\begin{eqnarray}
  \Delta \theta &\simeq& \frac{1}{12} \frac{\pi}{\Gamma(\frac{1}{6})
  \Gamma(\frac{13}{6})} \left(\!\frac{2}{3}\!\right)^{\frac{1}{3}}
  {\epsilon}^2 \alpha ({m_\epsilon} L)
  \left( \frac{{m_\epsilon}}{\omega}\right)^{\frac{1}{3}}
  \left( \frac{{\epsilon} e{B}}{{m_\epsilon}^2}
  \right)^{\frac{2}{3}}, \nonumber\\
  &&\quad \text{for}\,\, \frac{3}{2}
  \frac{\omega}{m_{\epsilon}} \frac{\epsilon e B}{m_\epsilon^2}\gg 1,
\label{eq:rotMCP}
\end{eqnarray}
which is valid above threshold and for a high number of allowed MCP Landau
levels. Similar formulas exist for ellipticity or the case of spin-0 MCPs
\cite{Ahlers:2006iz}. Note that this rotation appears to become independent of
$m_\epsilon$ in the small-mass limit. In practice, once the associated Compton
wavelength $\sim 1/m_\epsilon$ becomes larger than the size of the strong
field, the field size acts as a cutoff reducing the effect. Precise
predictions then require computations of polarization tensors in inhomogeneous
fields which is a challenge for standard methods and remains an interesting
question for future research.

\section{New-physics sensitivity scales of strong-field experiments}

In order to put the capabilities of the particle-physics potential of optical
experiments with high-intensity lasers into a greater context, let us draw a
comparison with other currently performed optical experiments, such as PVLAS
\cite{Zavattini:2007ee}, BMV \cite{Robilliard:2007bq}, ALPS
\cite{Ehret:2007cm}, LIPSS \cite{Afanasev:2006cv}, OSQAR \cite{OSQAR}, GammeV
\cite{Chou:2007zzc,Chou:2008gr}. In all these experiments, optical probe
lasers traverse a magnetic field of $\mathcal O(1-10 \, \text{Tesla})$ and
length $\mathcal O(1-10 \, \text{m})$. Whereas the reachable field strengths
are comparatively small, e.g., if measured in units of the QED critical field
strength of $B_{\text{cr}}\simeq 4 \times 10^9$ Tesla, the length of the interaction
region is macroscopic. As already discussed above in the context of QED
birefringence, the latter can even be enhanced by placing the field into a
high-finesse cavity such that the signal is increased by a factor
$N_{\text{pass}}$ counting the number of passes of the probe laser inside the
cavity.

By contrast, high-intensity laser systems provide for an interaction region
only of the order of $\mathcal{O}(10-100 \, \mu\text{m})$; also cavities are
of no use, since pulse durations on the femtosecond scale are far too short
compared to the time scale for a multiple pass. Nevertheless, the extreme
intensity can compensate for these disadvantages. In the following, we base
our estimates on a reference scenario with a peak intensity of
$I=10^{27}$Watt/cm$^2$ and a laser focus spot size of $L\simeq 50\mu$m
\cite{GHK}. Although these are optimistic values even for ELI, the
technological developments are rather rapid these days, such that these
parameters may be reliably accessible by the time when ELI is operating in a
stable manner. Note that recent ideas on higher-harmonic focusing by
oscillating plasma mirrors or schemes based on relativistically flying mirrors
might lead to even higher intensities \cite{Pukhov}.

Let us first concentrate on ALP scenarios, focusing on a parameter range
satisfying $Lm_\phi^2/\omega\ll 1$ (sub-eV particle masses). Then the relevant
combined dimensionless parameter is $g B L$. For instance for PVLAS, this
parameter is $gBL|_{\text{PVLAS}}\simeq 5 \times (g/\text{GeV}^{-1})$; the
other magnet-based experiments mentioned above lie in a similar ball park. As
a result, typical bounds on the coupling $g$, (e.g., resulting from the
nonobservation of photon regeneration behind a wall) are in the range of
$g\lesssim 10^{-5} \dots 10^{-6}\text{GeV}^{-1}$ for sub-eV masses
$m_\phi$. The corresponding ELI parameters are expected to give
\begin{equation}
gBL=3.3\times10^3 \, \frac{g}{[\text{GeV}^{-1}]}\,
\sqrt{\frac{I}{[10^{27} \frac{\text{Watt}}{\text{cm}^2}]}}\,  \frac{L}{[50 \,\mu\text{m}]},
\label{eq:ALPELI}
\end{equation}
yielding a prefactor which exceeds that of magnet-based optical experiments by
up to 3 orders of magnitude.

However, this improvement does not directly translate into a comparable
increase of sensitivity, due to the lack of cavity enhancements and the
necessity of pulse-probe synchronization. As an example estimate, let us
consider a regeneration or afterglow experiment \eqref{regALPps}, using a
probe laser of the Peta-Watt class delivering $\sim 10^{21}$ photons per
shot. Alternatively, a fraction of the ELI beam could be coupled out of the
beam and successively be used as a probe beam. The latter setup would
also be advantageous for issues of pulse synchronization. 

Assuming single-photon detection per pulse behind the wall or in the
afterglow, the sensitivity range for the ALP-photon coupling $g$ yields,
\begin{eqnarray}
\frac{g}{[\text{GeV}^{-1}]}& \gtrsim &3.4\times 10^{-9}\, \sqrt{\frac{[10^{27}
    \frac{\text{Watt}}{\text{cm}^2}]}{I}}\,  \frac{[50 \,\mu\text{m}]}{L}
\nonumber\\
&& \times
\sqrt[4]{\frac{10^{21}[\text{pulse}^{-1}]}{n_{\text{in}}}}\label{eq:ELIsensALP}
\end{eqnarray}
for ALP masses in the sub-eV range. This should be compared with the current
best laboratory bounds excluding ALP couplings of $g\gtrsim 10^{-6}$
GeV${}^{-1}$ or chameleonic couplings $g\gtrsim 2.5\times 10^{-7}$
GeV${}^{-1}$. Also ALP rotation and ellipticity signals could be enhanced in
comparison with standard optical experiments, but the potential improvement of
ALP parameter bounds might not be as dramatic as from regeneration or
afterglow experiments.  In any case, we conclude that ELI has the potential to
significantly improve existing laboratory bounds for Standard Model extensions
involving ALPs.

Let us turn to the MCP case. From \Eqref{eq:rotMCP}, we deduce that ELI
may yield the following maximum rotation $\delta \theta$ of the polarization
axis of a probe beam (at $\theta=\pi/2$):
\begin{equation}
\delta\theta
=4.1\times 10^8 \epsilon^{8/3}\,
\left(\frac{I}{[10^{27} \frac{\text{Watt}}{\text{cm}^2}]}\right)^{4/3} 
\left(\frac{\text{eV}}{\omega}\right)^{1/3}  \frac{L}{[50 \,\mu\text{m}]}.
\end{equation}
Assuming a detection sensitivity of $\delta \theta|_{\text{sens}}\simeq 10
\text{nrad}$, ELI will be sensitive to minicharge couplings down to
$\epsilon\gtrsim \mathcal{O}(10^{-7})$ for optical probe lasers and sub-eV MCP
masses. This is of the same order of magnitude as the current best laboratory
bounds from PVLAS \cite{Zavattini:2007ee,Ahlers:2007qf} and in the same ball park as
cosmological observations \cite{Melchiorri:2007sq}.

\section{Conclusions}

The prospect of high-intensity laser systems being currently worldwide under
intense development is a strong motivation for reconsidering aspects of
fundamental physics in strong fields. This subject had early been initiated
even before the full advent of quantum field theory. In particular, the
long-standing prediction of quantum-induced nonlinear self-interactions of
macroscopic magnetic fields by Heisenberg, Euler and others is still awaiting
its experimental verification. High-intensity laser systems are good candidates
for completing this quest.

In addition to confirming our expectations about nonlinearities induced by
known degrees of freedom of the Standard Model of particle physics,
strong-field experiments have recently proved very useful to explore new
regions in the parameter space of hypothetical new-physics degrees of
freedom. Strong fields in combination with optical probes have turned out to
be particularly sensitive to weakly coupled particles with light masses in the
sub-eV range. High-intensity lasers can add a new chapter to this story by
giving experimental access to unprecedented field-strength values. In this
work, we have argued that such laser systems may not only discover the
QED-induced nonlinearities for macroscopic fields for the first time, but also
search for unexpected optical signals that would point to new particle
candidates such as axion-like particles (ALPs) or minicharged particles
(MCPs). In particular for ALPs, optical probing of strong fields based on the
light-shining-through-wall scheme or the afterglow mechanism can exceed
current bounds by 2 or 3 orders of magnitude. 

It is worthwhile to emphasize that optical experiments typically
test a regime characterized by momentum transfers below the eV scale. This
clearly distinguishes them from experiments looking for astrophysical bounds.
Nevertheless, astrophysical observations in combination with energy-loss
arguments impose strong constraints, e.g., on the ALP coupling $g\lesssim
10^{-10}$GeV${}^{-1}$ for ALP masses in the eV range and below
\cite{Raffelt:2006cw}, or on MCP couplings $\epsilon\leq 2\times 10^{-14}$ for
$m_{\epsilon}$ below a few keV \cite{Davidson:2000hf}. However, since the
underlying solar physics involves keV momentum transfer scales, these bounds
apply to laboratory transfer scales ($\sim \mu$eV) only if the coupling values
are extrapolated over these many orders of magnitude \cite{Jaeckel:2006xm}. It
is precisely this assumption which has been put into question by various
models \cite{Masso:2006gc,Mohapatra:2006pv,Jain:2006ki,Foot:2007cq,cham,%
  Antoniadis:2007sp} and which can be checked or falsified by a particle
discovery at strong-field experiments such as POLARIS or ELI.  Indeed, current
strong-field laboratory experiments begin to enter the parameter regime which
has previously been accessible only to cosmological and astrophysical
considerations \cite{Ahlers:2007qf}.

For the special case of scalar ALPs, it has been pointed out
\cite{Dupays:2006dp,Adelberger:2006dh} that the scalar-ALP-photon coupling can
be severely constrained by direct searches for non-Newtonian forces
\cite{Kapner:2006si}. For instance, in simple models the bounds on the allowed
parameter region can reach up to $g\lesssim 1.6\times
10^{-17}\text{GeV}^{-1}$. (The bounds on chameleonic ALP couplings are somewhat
relaxed owing to the skin-depth effect \cite{cham}, but a significant
parameter range is excluded). Since gravitational or fifth-force
experiments measure ALP-matter interactions, the bounds rely on implicit
assumptions on additional ALP-matter couplings. Typically these interactions
are assumed to be generated by fluctuations from the ALP-photon
coupling. However, if also microscopic ALP-matter couplings are present, these
experiments only measure a sum of both microscopic and photon-induced ALP-matter
interactions, the combination of which could be accidentally small. Optical
measurements in high-intensity laser systems can therefore complement these
bounds and exclude this loophole, as they are directly sensitive to the ALP-photon
interaction.

In case of a positive signal, high-intensity lasers could not only discover a
new particle but also contribute to the particle's identification. Whereas the
field strength, length and frequency dependence can distinguish between ALPs or
MCPs, the signs of ellipticity and rotation are characteristic for spin and
parity \cite{Ahlers:2006iz}. Light-shining-through-wall or afterglow
experiments are indicative for additional matter couplings. Further
experiments have been suggested such as Schwinger-type MCP pair production
\cite{Gies:2006hv} or hidden-photon searches \cite{Ahlers:2007rd} which may
also become realizable at ELI.

It should nevertheless be stressed once more that all the above estimates are
based on various approximations. In particular, the homogeneous-field
assumption is questionable as the typical variation scale of the strong field
can be of the same order of magnitude as the new particle's Compton
wavelength. More detailed theoretical analyses are certainly required for
precise estimates, and new unknown and surprising effects may arise from this
interesting equality of scales.

\begin{acknowledgement}
  The author acknowledges collaboration and/or interesting discussions with
  E.G.~Adelberger, M.~Ahlers, R.~Alkofer, A.~Di Piazza, W.~Dittrich,
  B.~D\"obrich, G.V.~Dunne, F.~Hebenstreit, T.~Heinzl, J.~Jaeckel, D.~Mota,
  R.~Sch\"utzhold, G.~Shaw, J.~Redondo, A.~Ringwald, and T.~Tajima on the topics
  presented here.  He is grateful to the organizers of the ELI workshop for
  providing such an excellent working environment and acknowledges the
  support by the European Commission under contract ELI pp 212105 in the
  framework of the program FP7 Infrastructures-2007-1. This work was supported
  by the DFG through grants Gi 328/1-3, Gi 328/5-1 and the SFB TR18.
\end{acknowledgement}

%
%

\end{document}